\documentclass[aps,prb,twocolumn,groupedaddress,showpacs]{revtex4}
\usepackage{graphicx}
\usepackage{dcolumn}
\usepackage{bm}
\usepackage{amsmath}

\begin{document}

%\preprint{}                      

\title{Equivalence of Gutzwiller and slave-boson mean-field theories\\
 for multi-band Hubbard models} 
 
\author{J.~B\"unemann and F.~Gebhard}
\affiliation{Fachbereich Physik, Philipps--Universit\"at Marburg, 
D--35032 Marburg, Germany}

\begin{abstract}% 
We demonstrate that a recently introduced slave-boson mean-field theory 
 is equivalent to our Gutzwiller theory for multi-band Hubbard models with general onsite 
interactions. We relate the different objects that appear in both approaches at zero temperature
 and discuss the limitations of both methods.

\end{abstract}
          
\maketitle 
             
\section{Introduction} \label{sec1}
\label{intro} 
Over the last few years the Gutzwiller variational theory has developed
into a useful tool for correlated multi-band 
systems\cite{bgw,prb98,epl,narlikar,katipaper,claudius,fabrizio1,fabrizio2,hofstaetter}.
Gutzwiller introduced his wave function to study 
ferromagnetism in the one-band Hubbard model\cite{gutzwiller}. 
For the evaluation of expectation values he used a classical counting scheme, 
the so-called 'Gutzwiller-approximation'; 
see also Ref.~[\onlinecite{vollhardtrmp}] 
for the discussion of its physical content and Ref.~[\onlinecite{buenemann}]
for its mathematical formulation. 
 It was found later that this somewhat 
ad hoc approximation is equivalent to an exact evaluation of expectation 
values in the limit of infinite spatial dimensions or lattice 
coordination number\cite{metzner,gebhard}.
 An evaluation of Gutzwiller wave-functions in this limit will be denoted 
the `Gutzwiller-theory' throughout this work. 
The limit of infinite spatial dimensions is also the central assumption 
in the dynamical mean-field 
theory (DMFT)\cite{vollhardt}. The solution of the
DMFT equations is quite 
challenging even for the one-band Hubbard model. In contrast,
 the corresponding Gutzwiller variational space
  is just one-dimensional and its minimisation is a trivial 
numerical task. 
Of course, the study of the electronic properties of real materials
requires the treatment of multi-band Hubbard models. This task was
accomplished some years ago in Refs.~[\onlinecite{bgw,prb98,narlikar}]. 

An alternative scheme 
to derive the Gutzwiller energy functional for a single band
is based on the slave-boson mean-field theory (SBMFT) of Kotliar and 
Rucken\-stein\cite{kotliar}. A generalization
of this approach which reproduces the results of the multi-band Gutzwiller
theory is straightforward for systems with only 
density-density interaction\cite{bgw,hasegawa,kotliar2}. 
It was only recently, however, that Lechermann et al.\cite{lechermann} 
succeeded to develop a SBMFT scheme that allows to investigate systems 
with general multi-band interactions. 
The authors interpret their approach as a generalization
 of the Gutzwiller theory. In fact, as we will show in this 
 work, both theories are completely equivalent.

Our article is structured as follows.
In Sect.~\ref{sec2} we introduce the multi-band Hamiltonian and the 
general class of Gutzwiller wave-functions. 
The equivalence of the SBMFT, as derived in Ref.~[\onlinecite{lechermann}], 
and the  Gutzwiller theory is demonstrated in Sect.~\ref{sec3}.
Finally, we critically discuss the limitations 
of both methods in Sect.~\ref{sec4}.  
         
\section{Hubbard models and Gutzwiller wave-functions}\label{sec2}
We investigate multi-band Hubbard models, described by 
the general class of Hamiltonians
\begin{equation}\label{1.1}
\hat{H}=\sum_{i \ne j;\sigma,\sigma'}t_{i,j}^{\sigma,\sigma'}\hat{c}_{i,\sigma}^{\dagger}\hat{c}_{j,\sigma'}^{\phantom{+}}
+\sum_i \hat{H}_{{\rm loc},i}=\hat{H}_0+\hat{H}_{\rm loc}\;.
\end{equation}
 Here, the first term describes the hopping of electrons between spin-orbital 
states 
 $\sigma,\sigma'$ on lattice sites $i,j$, respectively. The Hamiltonian 
$\hat{H}_{{\rm loc},i}$ contains all local terms, i.e., the two-particle Coulomb 
interactions and the orbital onsite-energies. For any lattice site $i$ 
one introduces the Fock-states $|I\rangle_{i}$, in which certain sets of 
spin-orbital states $\sigma$  are occupied\cite{prb98,narlikar}. These 
states form a basis of the local atomic Hilbert space and can be used to 
write any other local multiplet state as 
\begin{equation}\label{1.2}
|\Gamma \rangle_{i}=\sum_{I}T^{(i)}_{I,\Gamma}|I\rangle_{i}\;.
\end{equation}
The most general Ansatz for a multi-band Gutzwiller wave-function has the form
\begin{equation}\label{1.3}
|\Psi_{\rm G}\rangle=\hat{P}_{\rm G}|\Psi_0\rangle=\prod_{i}\hat{P}_{i}|\Psi_0\rangle\;,
\end{equation}
where $|\Psi_0\rangle$ is a normalized single-particle product state and the 
local Gutzwiller correlator is defined as 
\begin{equation}\label{1.4}
\hat{P}_{i}=\sum_{\Gamma,\Gamma^{\prime}}\lambda^{(i)}_{\Gamma,\Gamma^{\prime}}
|\Gamma \rangle_{i} {}_{i}\langle \Gamma^{\prime} |\;.
 \end{equation}
Here, the states $|\Gamma \rangle_{i}$ can be an arbitrary atomic basis and the 
numbers $\lambda^{(i)}_{\Gamma,\Gamma^{\prime}}$ are variational parameters. In our 
first work, Ref.~[\onlinecite{prb98}], we assumed that the multiplets 
$|\Gamma \rangle_{i}$ are 
the eigenstates of $\hat{H}_{{\rm loc},i}$ and  
$\lambda^{(i)}_{\Gamma,\Gamma^{\prime}}\sim \delta_{\Gamma,\Gamma^{\prime}}$. The more 
general Ansatz~(\ref{1.4}) was first evaluated in Ref.~[\onlinecite{narlikar}] 
for Hermitian operators $\hat{P}_{{\rm G},i}$. The non-Hermitian case has been 
studied in Ref.~[\onlinecite{fabrizio2}]. 
In the following, we drop the site index when we deal with purely local 
quantities. 

In general, the uncorrelated local density matrix 
\begin{equation}\label{1.8}
C^{0}_{\sigma,\sigma'}\equiv \langle  \hat{c}^{\dagger}_{\sigma} \hat{c}^{}_{\sigma'}\rangle_{\Psi_0}
\end{equation}
is not diagonal. It is then useful to introduce a second orbital basis, 
defined by the operators $\hat{h}^{(\dagger)}_{\gamma}$, which, 
by construction, have a diagonal local density matrix,
\begin{equation}\label{hop}
\hat{h}^{(\dagger)}_{\gamma}
=\sum_{\sigma}u^{(*)}_{\gamma,\sigma}\hat{c}^{(\dagger)}_{\sigma}\;,
\;
\langle  \hat{h}^{\dagger}_{\gamma} \hat{h}^{}_{\gamma'}\rangle_{\Psi_0}
=\delta_{\gamma,\gamma'}n^0_{\gamma} \; .
\end{equation}
Within the Gutzwiller theory one usually works
in the new `$h$-representation' because all formulae
have a much simpler form than in the original `$c$-representation'. 
However, in order to show the equivalence of the Gutzwiller theory 
with the slave-boson results in section~\ref{sec3}, we  
have to work with both representations simultaneously. The operators 
$\hat{h}^{\dagger}_{\gamma}$ define Fock states $|H\rangle$ 
which can also be used to write the multiplet states (\ref{1.2}) as
\begin{equation}
|\Gamma \rangle=\sum_{H}T_{H,\Gamma}|H\rangle\;,
\end{equation}
where the coefficients $T_{H,\Gamma}$ and $T_{I,\Gamma}$ are related through
\begin{equation}
T_{H,\Gamma}=\sum_{I}\Omega_{H,I}T_{I,\Gamma}\;\;,\;\;
\Omega_{H,I}\equiv\langle H | I \rangle \;.
\end{equation}

\section{Comparison}\label{sec3}
The calculation of expectation values for Gutzwiller wave-functions 
is a straightforward task, once the basic structure of diagrams 
 in infinite dimensions is under\-stood\cite{bgw,prb98,narlikar}. 
In contrast, the derivation 
of the generalized 
slave-boson mean-field theory in Ref.~[\onlinecite{lechermann}] 
requires a number of subtle ideas. 
Furthermore, there is not a clear correspondence of all the mathematical 
objects that appear in both approaches. Therefore, we are not going 
to compare any  particular steps of the two quite different derivations, 
but focus on the  final energy functional at zero temperature in order 
 to show the equivalence of both approaches. For all details of the 
derivations we refer the reader to Refs.~[\onlinecite{bgw,prb98,narlikar,lechermann}]. 

\subsection{Local energy}

In infinite dimensions, the expectation value of the local 
Hamiltonian $\hat{H}_{{\rm loc},i}$ in the Gutzwiller wave function reads
\begin{equation}\label{1.5}
\langle \hat{H}_{{\rm loc},i} \rangle_{\Psi_{\rm G}}=\sum_{\Gamma_{1}\ldots\Gamma_{4}}\lambda_{\Gamma_{2},\Gamma_{1}}^{*}\lambda_{\Gamma_{3},\Gamma_{4}}^{}
E^{\rm loc}_{\Gamma_{2},\Gamma_{3}}m^0_{\Gamma_1,\Gamma_4}
\;,
\end{equation}
where 
\begin{eqnarray}\label{1.6}
E^{\rm loc}_{\Gamma_{2},\Gamma_{3}}&\equiv&\langle 
\Gamma_{2} | \hat{H}_{{\rm loc},i}  | \Gamma_{3}\rangle\;,\\
\label{1.6b}
m^0_{\Gamma_1,\Gamma_4} &\equiv& \langle \left( |\Gamma_{1} \rangle   \langle  \Gamma_{4} |\right)  \rangle_{\Psi_0}\;.
\end{eqnarray}
The expectation value (\ref{1.6b}) can be written as
\begin{eqnarray}\label{1.7b}
m^0_{\Gamma_1,\Gamma_4}
&=&\sum_{H_1,H_4}T_{H_1,\Gamma_1}^{} T_{H_4,\Gamma_4}^{*}m^{0}_{H_1,H_4}\;.\\\label{1.7}
&=&\sum_{H}T_{H,\Gamma_1}^{}T_{H,\Gamma_4}^{*}m^{0}_{H}
\end{eqnarray}
because, for a diagonal local density-matrix in the $h$-representation, one readily finds 
\begin{eqnarray}\label{1.9}
m^{0}_{H,H'}&=&\delta_{H,H'}m^{0}_{H}\;,\\\label{1.9bb}
m^{0}_{H}&\equiv&\prod_{\gamma({\rm occ.})}n^0_{\gamma} \prod_{\gamma'({\rm unocc.})}(1-n^0_{\gamma'})\;.
\end{eqnarray}
In order to make contact with the results in Ref.~[\onlinecite{lechermann}] we 
 need to bring (\ref{1.7}) in the slightly more complicated form
\begin{equation}\label{1234}
m^0_{\Gamma_1,\Gamma_4}=\sum_{H,H',I}
T_{H,\Gamma_1}^{}\Omega^{*}_{H,I}\sqrt{m^{0}_{H}}
T_{H',\Gamma_4}^{*}\Omega^{}_{H',I}\sqrt{m^{0}_{H'}}
\end{equation}
which is equivalent to (\ref{1.7}) because of the completeness relation
\begin{equation}\label{ort}
\sum_{I}\Omega_{H',I}\Omega^{*}_{H,I}=\delta_{H,H'}\;.
\end{equation}
We now introduce the new variational parameters 
\begin{equation}\label{phi}
\varphi_{\Gamma,I}\equiv\sum_{\Gamma',H}\lambda_{\Gamma,\Gamma'}
T_{H,\Gamma'}^{*}\Omega^{}_{H,I}\sqrt{m^{0}_{H}}
\end{equation}
which allow us to write the expectation value (\ref{1.5}) as
 \begin{equation}\label{2.1}
\langle \hat{H}_{{\rm loc},i}\rangle_{\Psi_{\rm G}}=\sum_{\Gamma,\Gamma'}\sum_{I}
\varphi^{*}_{\Gamma,I}\varphi^{}_{\Gamma',I}E_{\Gamma,\Gamma'}^{\rm loc}
\end{equation}
This equation has exactly the same form as equation (\underline{47}) 
 in Ref.~[\onlinecite{lechermann}], after the slave-boson operators
 $\phi_{\Gamma,I}$ have been replaced by their mean-field expectation values,
$ \phi_{\Gamma,I}\mapsto \varphi_{\Gamma,I}$.

\subsection{Local constraints}
The variational parameters need to obey certain constraints which naturally arise in 
the evaluation in infinite dimensions. These are
\begin{eqnarray}\label{1.10a}
\langle\hat{P}^{\dagger}\hat{P}^{}\rangle_{\Psi_0}&=&1\;,\\\label{1.10b}
\langle  \hat{c}^{\dagger}_{\sigma} 
\hat{c}^{}_{\sigma'}\hat{P}^{\dagger}\hat{P}^{} \rangle_{\Psi_0}&=&\langle
 \hat{c}^{\dagger}_{\sigma}\hat{c}^{}_{\sigma'}   \rangle_{\Psi_0}\;. 
\end{eqnarray}
Note  that moving the operator $\hat{P}^{\dagger}\hat{P}^{}$  relative to 
$\hat{c}^{\dagger}_{\sigma}$ or $\hat{c}^{}_{\sigma'}$ 
 in (\ref{1.10b})  would not alter the whole 
set of constraints. A set 
 of constraints equivalent to (\ref{1.10b}) is obtained when we use the 
operators $\hat{h}^{(\dagger)}_{\gamma} $,
\begin{equation}\label{1.10c}
\langle \hat{h}^{\dagger}_{\gamma} 
\hat{h}^{}_{\gamma'}\hat{P}^{\dagger}\hat{P}^{}  \rangle_{\Psi_0}=\langle
 \hat{h}^{\dagger}_{\gamma}\hat{h}^{}_{\gamma'}   \rangle_{\Psi_0}\;. 
\end{equation}
The constraint (\ref{1.10a}) can be written as 
\begin{equation}\label{5.5}
\sum_{\Gamma,\Gamma_1,\Gamma_2}
\lambda_{\Gamma,\Gamma_1}^{*}\lambda_{\Gamma,\Gamma_2}^{}
m^{0}_{\Gamma_1,\Gamma_2}=1\;,
\end{equation}
which, by use of eqs.\ (\ref{1234}) and (\ref{phi}), is found to be 
  equivalent to 
\begin{equation}\label{nb1}
\sum_{\Gamma,I}\varphi^{*}_{\Gamma,I}\varphi^{}_{\Gamma,I}=1\;.
\end{equation}
This is equation (\underline{28}) in Ref.~[\onlinecite{lechermann}] at 
 mean-field level.
For the constraints (\ref{1.10c}) it follows that
\begin{eqnarray} \label{2345}
\sum_{\Gamma,\Gamma_1,\Gamma_2}\sum_{H_1,H_2}&&
\lambda_{\Gamma,\Gamma_1}^{*}\lambda_{\Gamma,\Gamma_2}^{}
T_{H_1,\Gamma_1}T_{H_2,\Gamma_2}^{*}\\\nonumber
&&\times \left \langle \left(\hat{h}^{\dagger}_{\gamma}
\hat{h}^{}_{\gamma'} |H_1 \rangle   \langle  H_2|\right) \right \rangle_{\Psi_0}=\langle\hat{h}^{\dagger}_{\gamma}\hat{h}^{}_{\gamma'}   \rangle_{\Psi_0}\;,
\end{eqnarray} 
where, due to eqs.\ (\ref{1.9}) and (\ref{1.9bb}), the expectation value 
on the l.h.s.\ can be written as
\begin{equation} \label{1.12}
\left \langle \left(\hat{h}^{\dagger}_{\gamma}
\hat{h}^{}_{\gamma'} |H_1 \rangle   \langle  H_2|\right) \right \rangle_{\Psi_0}
=\langle H_2 | \hat{h}^{\dagger}_{\gamma}
\hat{h}^{}_{\gamma'}  | H_1 \rangle
\sqrt{m_{H_1}^0 m_{H_2}^0} \;.
\end{equation}
Then, the identity  
\begin{equation}
\langle H_2 | \hat{h}^{\dagger}_{\gamma}
\hat{h}^{}_{\gamma'}  | H_1 \rangle=
\sum_{I_1,I_2}\Omega^{}_{H_2,I_2} 
\langle I_2 | \hat{h}^{\dagger}_{\gamma}
\hat{h}^{}_{\gamma'}  | I_1 \rangle \Omega^{*}_{H_1,I_1} 
\end{equation}
transforms eq.\ (\ref{2345}) into the form 
\begin{equation}\label{nb2a}
\sum_{\Gamma}\sum_{I,I'}
\varphi^{*}_{\Gamma,I}\varphi_{\Gamma,I'}
\langle I' |\hat{h}^{\dagger}_{\gamma}\hat{h}^{}_{\gamma'} | I \rangle
= \langle  \hat{h}^{\dagger}_{\gamma}\hat{h}^{}_{\gamma'}\rangle_{\Psi_0}\;.
\end{equation}
These equations can be transformed 
to the $c$-re\-pre\-senta\-tion which leads to
\begin{equation}\label{nb2}
\sum_{\Gamma}\sum_{I,I'}
\varphi^{*}_{\Gamma,I'}\varphi_{\Gamma,I}
\langle I |\hat{c}^{\dagger}_{\sigma}\hat{c}^{}_{\sigma'} | I' \rangle
= \langle  \hat{c}^{\dagger}_{\sigma}\hat{c}^{}_{\sigma'}\rangle_{\Psi_0}\;.
\end{equation}
Equation (\ref{nb2}) is equivalent to equation (\underline{29}) 
 in Ref.~[\onlinecite{lechermann}] at mean-field level.
  
\subsection{Hopping renormalization}
Finally, we investigate the expectation value of the electron transfer 
operators in the Hamiltonian (\ref{1.1}). 
In infinite dimensions one finds that such an expectation value has the form ($i\neq j$)
\begin{equation}\label{1.13a}
\langle  \hat{c}_{i,\sigma_1}^{\dagger}\hat{c}_{j,\sigma_2}^{\phantom{+}} \rangle_{\Psi_{\rm G}}=\sum_{\sigma'_1,\sigma'_2}r_{\sigma_1}^{\sigma'_1}\left( r_{\sigma_2}^{\sigma'_2}\right)^{*}\langle  
\hat{c}_{i,\sigma'_1}^{\dagger}\hat{c}_{j,\sigma'_2}^{\phantom{+}} \rangle_{\Psi_{0}}\;,
\end{equation}
or, alternatively, in the $h$-representation
\begin{equation}\label{1.13}
\langle  \hat{h}_{i,\gamma_1}^{\dagger}\hat{h}_{j,\gamma_2}^{\phantom{+}} \rangle_{\Psi_{\rm G}}=\sum_{\gamma'_1,\gamma'_2}q_{\gamma_1}^{\gamma'_1}\left( q_{\gamma_2}^{\gamma'_2}\right)^{*}\langle  
\hat{h}_{i,\gamma'_1}^{\dagger}\hat{h}_{j,\gamma'_2}^{\phantom{+}} \rangle_{\Psi_{0}}\;.
\end{equation}
The local renormalization-matrix is most easily calculated in the 
$h$-representation\cite{fabrizio1,narlikar} where it has the rather simple form
\begin{equation}\label{1.14}
q_{\gamma}^{\gamma'}=\frac{1}{n^{0}_{\gamma'}}\langle\hat{P}^{\dagger}
\hat{h}_{\gamma}^{\dagger} \hat{P} 
\hat{h}_{\gamma'}^{}\rangle_{\Psi_{0}}\;.
\end{equation}
The matrix $r_{\sigma}^{\sigma'}$ in the $c$-representation can then 
be derived from $q_{\gamma}^{\gamma'}$ by the transformation 
\begin{equation}\label{trafo}
r_{\sigma}^{\sigma'}=\sum_{\gamma,\gamma'}q_{\gamma}^{\gamma'}u^{}_{\gamma,\sigma}
u^{*}_{\gamma',\sigma'}\;.
\end{equation}
With eqs.\ (\ref{1.3}) and (\ref{1.4}), the matrix $q_{\gamma}^{\gamma'}$ reads
 explicitly 
\begin{eqnarray}\label{1.15}
q_{\gamma}^{\gamma'}=\frac{1}{n^{0}_{\gamma'}}
\sum_{\Gamma_1\ldots\Gamma_4}\lambda^{*}_{\Gamma_2,\Gamma_1}
\lambda^{}_{\Gamma_3,\Gamma_4}
\langle \Gamma_2|
\hat{h}_{\gamma}^{\dagger}
|\Gamma_3\rangle&&\\ \nonumber
\times \left \langle
\left(|\Gamma_1  \rangle
\langle \Gamma_4 |  \hat{h}_{\gamma'}
  \right)
\right \rangle_{\Psi_0}&& \;,
\end{eqnarray}
where the expectation value in (\ref{1.15}) can be written as
\begin{eqnarray}\nonumber
\left\langle
\left(|\Gamma_1  \rangle
\langle \Gamma_4 |  \hat{h}_{\gamma'}
  \right)
\right \rangle_{\Psi_0}&=&\sum_{H_1,H_4}T_{H_1,\Gamma_1}T^{*}_{H_4,\Gamma_4}
\langle H_4 |\hat{h}_{\gamma'}^{}  | H_1 \rangle\\\label{1.16}
&&\times \sqrt{m_{H_1}^0 m_{H_4}^0} \sqrt{\frac{n^0_{\gamma'}}{1-n_{\gamma'}^0}}\;.
\end{eqnarray}
With  eqs.\ (\ref{ort}) and (\ref{phi}) we can rewrite (\ref{1.15}) as
\begin{eqnarray} \nonumber
q_{\gamma}^{\gamma'}&=&\sqrt{\frac{1}{n^0_{\gamma'}(1-n^0_{\gamma'})}}\sum_{\Gamma,\Gamma'}
\sum_{I,I'}\varphi^{*}_{\Gamma,I}\varphi^{}_{\Gamma',I'}\\\label{5.1}
&&\times\langle \Gamma|
\hat{h}_{\gamma}^{\dagger}
|\Gamma'\rangle
\langle I' |\hat{h}_{\gamma'}^{}  | I \rangle\;.
\end{eqnarray}
The transformation (\ref{trafo}) from the $h$-representation 
to the $c$-representation
is not as straightforward as the corresponding 
 transformation from eq.\ (\ref{nb2a}) to eq.\ (\ref{nb2}). 
Whereas the transformation with respect to the lower index $\gamma$
is still simple
 \begin{equation}
\langle \Gamma|
\hat{c}_{\sigma}^{\dagger}
|\Gamma'\rangle =\sum_{\gamma}  \langle \Gamma|
\hat{h}_{\gamma}^{\dagger}
|\Gamma'\rangle u_{\gamma,\sigma}\;,
 \end{equation}
for the upper index $\gamma'$ we need to take into account the factor 
 $\sqrt{1/(n^0_{\gamma'}(1-n^0_{\gamma'}))}$ in (\ref{5.1})
 which also depends on $\gamma'$. For this purpose,
we introduce the hole density-matrix $\tilde{D}$ with the elements  
\begin{equation}\label{1.8b}
D^{0}_{\sigma,\sigma'}\equiv \langle  \hat{c}^{}_{\sigma'} 
\hat{c}^{\dagger}_{\sigma}\rangle_{\Psi_0}\;,
\end{equation}
in addition to the density-matrix $\tilde{C}$ already defined in (\ref{1.8}).
Then the transformation (\ref{trafo}) for the upper index $\gamma'$ 
can be carried out along the lines
\begin{eqnarray} 
\sum_{\gamma'}u^{*}_{\gamma',\sigma'}\frac{\hat{h}^{}_{\gamma'}}
{\sqrt{n^0_{\gamma'}(1-n^0_{\gamma'})}}&=&\sum_{\gamma',\tilde{\sigma}}
\frac{u^{*}_{\gamma',\sigma'}u^{}_{\gamma',\tilde{\sigma}}}
{\sqrt{n^0_{\gamma'}(1-n^0_{\gamma'})}}\hat{c}^{}_{\tilde{\sigma}}\\ \nonumber
&\equiv&\sum_{\tilde{\sigma}} 
\left (( \tilde{C}^{0} \tilde{D}^{0})^{-\frac{1}{2}} \right)_{\tilde{\sigma},\sigma'} \hat{c}^{}_{\tilde{\sigma}} \;.
\end{eqnarray}
Here, we used the notation 
\begin{equation}
\sum_{\gamma'}\label{678}
\frac{u^{*}_{\gamma',\sigma'}u^{}_{\gamma',\tilde{\sigma}}}
{\sqrt{n^0_{\gamma'}(1-n^0_{\gamma'})}}= 
  \left (( \tilde{C}^{0} \tilde{D}^{0})^{-\frac{1}{2}} \right)_{\tilde{\sigma},\sigma'}\;.
\end{equation}
With eq.\ (\ref{678}) and with 
$\langle I' |\hat{c}_{\tilde{\sigma}}^{}  | I \rangle=\langle I |\hat{c}_{\tilde{\sigma}}^{\dagger}  | I' \rangle$  we can finally write the renormalization-matrix in the 
$c$-representation as
\begin{eqnarray} \nonumber
r_{\sigma}^{\sigma'}&=&\sum_{\Gamma,\Gamma'}
\sum_{I,I'}\varphi^{*}_{\Gamma,I}\varphi^{}_{\Gamma',I'}\langle \Gamma|
\hat{c}_{\sigma}^{\dagger}
|\Gamma'\rangle\\\label{5.2}
&&\times
\sum_{\tilde{\sigma}}
 \left (( \tilde{C}^{0} \tilde{D}^{0})^{-\frac{1}{2}} \right)_{\tilde{\sigma},\sigma'}
\langle I |\hat{c}_{\tilde{\sigma}}^{\dagger}  | I' \rangle\;.
\end{eqnarray}
This expression matches equation (\underline{37}) in Ref.~[\onlinecite{lechermann}]
 at mean field level, apart from the fact that there
 the constraints have been used to write the matrices 
$\tilde{C}^{0} \tilde{D}^{0}$ as a function of the fields 
$\varphi^{}_{\Gamma,I}$; see equations  
(\underline{35}) and (\underline{36}) in Ref.~[\onlinecite{lechermann}]. 
However, as long as the constraints are fulfilled, this makes no difference because
 it does not change the variational energy functional.

\section{Discussion}\label{sec4}
In this work we showed that the multi-band slave-boson 
mean-field theory of Lechermann et al.\ reproduces correctly 
the energy functional of the multi-band Gutzwiller theory developed earlier. 
As a byproduct we were able to show how the different objects that appear in both 
 approaches are related. This will turn out to be important   
 for the comparison of future numerical results. 

We believe that there 
 are good reasons to prefer the derivation based on Gutzwiller
 wave-functions over the slave-boson mean-field theory. 
First, these wave functions are well defined and 
they are evaluated exactly in the unambiguous limit of infinite 
 spatial dimensions ($D\to \infty$). Therefore, e.g., the inclusion of 
superconducting pair-correlations was straightforward\cite{narlikar,katipaper}.
 In contrast, the slave-boson mean-field 
 derivation is uncontrolled and quite adjustable in its outcome.  
  Of all the different equations that one may derive within such an 
  approach, the `right ones' are usually identified by some 
sophisticated guess. 
 This guess, not surprisingly, always turns out to be equivalent to 
the Gutzwiller theory.
 This very equivalence is, by far, 
the most convincing argument for the credibility of the  SBMFT results. 

The Gutzwiller theory can also be used to calculate quasi-particle 
excitations within a Fermi-liquid ap\-proach\cite{narlikar,thul} as well as 
 spin-wave excitations\cite{spinwave}. The quasi-particle bands
 in the Gutzwiller theory coincide with those derived in the SBMFT. 
 Therefore the zero-temperature spectral properties are equivalent 
 in both approaches.

The ground-state energy functional also provides the Landau parameters
 for the description of thermodynamic properties\cite{vollhardtrmp}. Therefore,
 both approaches are equivalent in the Fermi-liquid regime when the 
 temperatures $T$ is much smaller than the Fermi temperature $T_{\rm F}$.
 Although the SBMFT equations 
can also be solved for  $T\approx T_{\rm F}$ or even for $T \gg T_{\rm F}$, 
the approximation breaks down in this temperature regime\cite{florian}.

In principle, the SBMFT could be improved by computing fluctuations
 around the saddle point. To the best of our knowledge, however, this promise, 
although  often made, has never materialized in any convincing improvement 
of the results, not even for the one-band Hubbard model\cite{arrigoni}. 
In contrast, for Gutzwiller wave-functions it is possible to calculate 
systematically $1/D$ corrections for all physical 
quantities\cite{gebhard,thul}. Such calculations allow to estimate the 
accuracy of the results in infinite dimensions and 
to improve them, if necessary. 

It should be kept in mind that the Gutzwiller theory is based on
 rather simple variational many-body wave functions which could be 
improved in many directions. Despite its limitations, however, the 
Gutzwiller theory appears to provide a suitable description of the 
 quasi-particle bands in ferromagnetic Nickel\cite{epl,narlikar}.

\end{document}